\documentclass[aps,prb,twocolumn,superscriptaddress,showpacs]{revtex4}
\usepackage{graphicx}
\usepackage{dcolumn}
\usepackage{bm}
\usepackage{color}
\usepackage{float}
\usepackage{amsmath}
\usepackage{amssymb}
\newcommand{\pdag}{{\phantom{\dagger}}}
\newcommand{\eye}{\mathrm{i}}
\newcommand{\bk}{{\bf k}}

\begin{document}
\title{Non-local correlations in the orbital selective Mott
phase of a one dimensional multi-orbital Hubbard model}
\author{S. Li}
\affiliation{Department of Physics and Astronomy, The University of Tennessee, Knoxville, Tennessee 37996, USA}
\author{N. Kaushal}
\affiliation{Department of Physics and Astronomy, The University of Tennessee, Knoxville, Tennessee 37996, USA}
\author{Y. Wang}
\affiliation{Department of Physics and Astronomy, The University of Tennessee, Knoxville, Tennessee 37996, USA}
\author{Y. Tang}
\affiliation{Department of Physics, Virginia Tech, Blacksburg, Virginia 24061, USA}
\author{G. Alvarez} 
\affiliation{Center for Nanophase Materials Sciences, Oak Ridge National Laboratory, Oak Ridge, Tennessee 37831, USA}
\affiliation{Computer Science and Mathematics Division, Oak Ridge National Laboratory, Oak Ridge, Tennessee 37831, USA.}
\author{A. Nocera}
\affiliation{Center for Nanophase Materials Sciences, Oak Ridge National Laboratory, Oak Ridge, Tennessee 37831, USA}
\affiliation{Computer Science and Mathematics Division, Oak Ridge National Laboratory, Oak Ridge, Tennessee 37831, USA.}
\author{T. A. Maier}
\affiliation{Center for Nanophase Materials Sciences, Oak Ridge National Laboratory, Oak Ridge, Tennessee 37831, USA}
\affiliation{Computer Science and Mathematics Division, Oak Ridge National Laboratory, Oak Ridge, Tennessee 37831, USA.}
\author{E. Dagotto}
\affiliation{Department of Physics and Astronomy, The University of Tennessee, Knoxville, Tennessee 37996, USA}
\affiliation{Materials Science and Technology Division, Oak Ridge National Laboratory, Oak Ridge, Tennessee 37831, USA}
\author{S. Johnston}
\affiliation{Department of Physics and Astronomy, The University of Tennessee, Knoxville, Tennessee 37996, USA}

\begin{abstract}  
We study non-local correlations in a three-orbital Hubbard model defined on an extended one-dimensional chain 
using determinant quantum Monte Carlo and density matrix renormalization group methods. We focus on 
a parameter regime with robust Hund's coupling, which produces an orbital selective Mott phase (OSMP) at 
intermediate values of the Hubbard $U$, as well as an orbitally ordered ferromagnetic 
insulating state at stronger coupling. An examination of the orbital- and spin-correlation functions indicates that 
the orbital ordering occurs before the onset of magnetic correlations in this parameter regime as a function of 
temperature. In the OSMP, we find that the self-energy for the itinerant electrons is momentum dependent, 
indicating a degree of non-local correlations 
while the localized electrons have largely momentum independent self-energies. 
These non-local correlations also produce relative shifts of the hole-like and electron-like bands 
within our model. The overall momentum dependence 
of these quantities is strongly suppressed in the orbitally-ordered insulating phase.
\end{abstract}

\date{\today}
\maketitle

\section{Introduction}  
In recent years the scientific community renewed its interest in understanding the properties of 
multi-orbital Hubbard models, and this has been intensified by 
the discovery of the iron-based superconductors.\cite{JohnstonAP2010,StewartRMP2011,
DaiNaturePhys2012,vanRoekeghem} On a theoretical front, this is a 
challenging problem due to a lack of non-perturbative methods for treating multi-orbital 
Hubbard models at intermediate or strong couplings and on extended systems. 
Nevertheless, considerable progress has been made using mean-field-based approaches,
\cite{vanRoekeghem,YuPRL2013, FanfarilloPRB2015,GeorgesReview,Zhang,Knecht,WernerPRL2007,
LiebschPRL2005,deMediciPRL2009,deMediciPRB2005,MukherjeePRB2016,SemonPreprint,Ferrero} 
resulting in new concepts such as that of a Hund's metal \cite{Zhang,YinNatureMaterials2011,GeorgesReview,LanataPRB2013} 
and the orbital-selective Mott phase (OSMP).\cite{Anisimov2002,GeorgesReview} These concepts are 
central to our understanding the paradoxical appearance of both localized and 
itinerant characteristics in many multi-orbital systems \cite{deMediciPRL2014,MannellaReview} 
and bad metallic behavior in the presence of sizable electronic correlations.\cite{deMediciPRL2014} 

The most widely used numerical approach in this context 
is single-site multi-orbital dynamical mean-field theory (DMFT).\cite{Footnote1,Georges,vanRoekeghem}  
Generally speaking, DMFT maps the full lattice problem onto an impurity problem embedded in an 
effective medium, which approximates the electron dynamics on a larger length scale 
as a local renormalization. \cite{Georges} 
While this technique has had considerable success in addressing many aspects of the OSMP and 
other physics related to the multi-orbital problem,\cite{Knecht,LiebschPRL2005,ChanPRB2009,
LiebschPRB2004,BiermannPRL2005,deMediciPRL2009,deMediciPRB2005,WernerPRL2007,LiebschPRB2010,GregerPRL2013} it is unable to 
capture spatial fluctuations and non-local correlations encoded in the $k$-dependent 
self-energy $\Sigma({\bf k},\omega)$. This is a potential short coming 
as non-local correlations are known to have an impact in the case of the 
single-band Hubbard model.\cite{GullEPL2008,ParkPRL2008} 
It is therefore important to assess the 
importance of such non-local effects on multi-orbital properties such as the OSMP.

To date, most non-perturbative studies have used 
cluster DMFT or the dynamical cluster approximation (DCA); \cite{NomuraPRB2015,LeePRB2011,LeePRL2010,BeachPRB2011,LeoPRL2008,SemonPreprint} 
however, these techniques are typically limited to a handful of sites when multiple orbitals 
are included in the basis. This is due to technical issues related to each choice in impurity solver,  
such as the Fermion sign problem in the case of quantum Monte Carlo or the exponential growth of the 
Hilbert space in the case of exact diagonalization. As a result, these studies have only addressed 
short-range spatial fluctuations. One study of the OSMP has been carried out on a larger two-dimensional cluster 
using determinant quantum Monte Carlo (DQMC). In that case, however, the OSMP was imposed by the model by 
assuming that electrons in a subset of orbitals were localized as Ising spins.\cite{BouadimPRL2009}  
In light of these limitations it is desirable to find situations where multi-orbital physics can be  
modeled on extended clusters that support long-range spatial fluctuations and where the properties   
under study emerge from the underlying many-body physics of the model. 

In this regard, one dimensional (1D) models are quite promising. For example, two recent density matrix 
renormalization group (DMRG) studies have been 
carried out for an effective 1D three-orbital model representative of the iron-based superconductors. 
\cite{RinconPRL2014,RinconPRB2014} More recently, it was demonstrated that 
DQMC simulations for a simplified version of the same model can also be carried out to low temperatures 
due to a surprisingly mild Fermion sign problem.\cite{LiuPreprint2016} These observations 
open the doorway to non-perturbative studies of this model on extended clusters, 
thus granting access to the momentum-resolved self-energies and non-local correlations.  
1D studies along these lines are also directly relevant for the recently-discovered quasi-1D   
selenide Ba$_{1-x}$K$_x$Fe$_2$Se$_3$.\cite{CaronPRB2011,CaronPRB2012,Patel,Shaui,LuoPRB2013,LuoPRB2014} 
In this context, it is important to note that DMFT becomes more accurate in higher dimensions and 
therefore one expects its ability to describe multi-orbital Mott physics in 1D to be 
diminished. 

Motivated by these considerations, we examine the properties of a three-orbital Hubbard 
Hamiltonian on an extended 1D cluster using DQMC and DMRG, with a particular focus 
on its $k$-resolved self-energies and spectral properties. 
We thus gain explicit access to non-local correlations occurring on longer length scales than 
those  addressed in previous non-perturbative studies. In general, we find that the 
OSMP leads to a mixture of localized and itinerant bands, where the former are characterized by a 
localized (momentum-independent)  
self-energy while the latter exhibits significant non-local (momentum-dependent) correlations. 
This leads to a band-dependent shift of the underlying band structures. We also 
identify an insulating state driven by orbital ordering in a region of parameter space previously 
associated with an OSMP.\cite{RinconPRL2014,RinconPRB2014}  

\section{Methods}\label{Sec:Methods} 
\subsection{Model Hamiltonian}
We study a simplified three-orbital model defined on a 1D chain as 
introduced in Ref. \onlinecite{RinconPRL2014}.  
This model displays a rich variety of phases including block ferromagnetism, 
antiferromagnetism, Mott insulting phases, metallic and band insulating phases, 
and several distinct OSMPs.\cite{RinconPRL2014,RinconPRB2014,LiuPreprint2016}
The Hamiltonian is $H = H_\mathrm{0} + H_\mathrm{int}$, where 
\begin{equation}\label{Eq:H0}
H_\mathrm{0} = - \sum_{\substack{\langle i,j \rangle\\\sigma,\gamma,\gamma^\prime}} 
t^\pdag_{\gamma\gamma^\prime} c^\dagger_{i,\gamma,\sigma}c^\pdag_{j,\gamma^\prime,\sigma} 
 + \sum_{i,\sigma,\gamma} (\Delta_\gamma-\mu) \hat{n}_{{i},\gamma,\sigma}  
\end{equation}
contains the non-interacting terms of $H$, and 
\begin{equation}\nonumber
\begin{split}
H_\mathrm{int}&=U\sum_{i,\gamma} \hat{n}_{i,\gamma,\uparrow}\hat{n}_{i,\gamma,\downarrow} + 
\left(U^\prime - \frac{J}{2}\right) \sum_{\substack{i,\sigma,\sigma^\prime\\
\gamma < \gamma^\prime}} \hat{n}_{i,\gamma,\sigma}\hat{n}_{i,\gamma,\sigma^\prime}\\
&+ J \sum_{i,\gamma<\gamma^\prime} S^\mathrm{z}_{i,\gamma} S^\mathrm{z}_{i,\gamma^\prime}
\end{split}
\end{equation}
contains the on-site Hubbard and Hund's interaction terms. Here, $\langle \dots \rangle$ 
denotes a sum over nearest-neighbors, $c^\dagger_{i,\gamma,\sigma}$ 
$(c^\pdag_{i,\gamma,\sigma})$ creates (annihilates) a spin $\sigma$ electron in orbital $\gamma = 1,2,3$ 
on site $i$, $\Delta_\gamma$ are the on-site energies for each orbital, 
$S^\mathrm{z}_{i,\gamma}$ is the z-component of the spin operator ${\bf S}_{i,\gamma}$,  
and $\hat{n}^\pdag_{i,\gamma,\sigma} = c^\dagger_{i,\gamma,\sigma}c^\pdag_{i,\gamma,\sigma}$ is the 
particle number 
operator. The pair-hopping and spin-flip terms of the interaction have been neglected in order 
to manage the sign problem in the DQMC calculations.  
  
\begin{figure}
 \includegraphics[width=0.75\columnwidth]{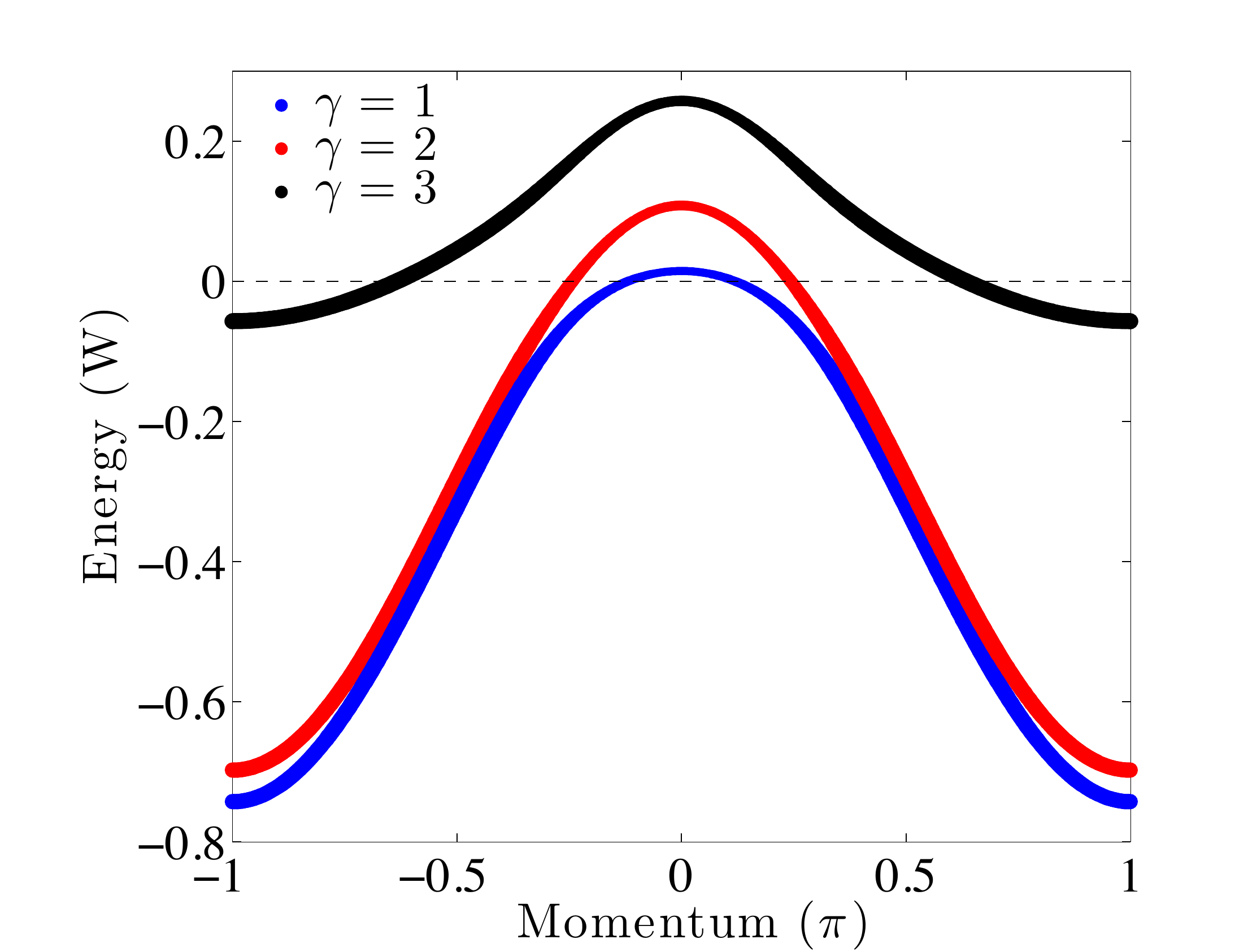}
 \caption{\label{Fig:Bands} (color online) A fat band plot of the non-interacting band structure 
 at a total filling of $\langle \hat{n} \rangle = 4$, where the thickness of the 
 lines indicates the majority orbital content of the band.  
 The top most band has the narrowest bandwidth and is primarily of orbital 3 character. 
 The lower two bands disperse over a much larger energy range and are primarily composed 
 of orbitals 1 and 2, respectively. }
\end{figure}

Following Ref. \onlinecite{RinconPRL2014}, we set $t_{11} = t_{22} = -0.5$, $t_{33} = -0.15$, 
$t_{13} = t_{23} = 0.1$, $t_{12} = 0$, $\Delta_1 = -0.1$, $\Delta_2 = 0$, and $\Delta_3 = 0.8$ 
in units of eV while the chemical potential $\mu$ is adjusted to obtain the desired filling. 
These parameters produce a non-interacting band structure   
analogous to the iron-based superconductors, with two hole-like bands  
centered at $k = 0$ and an electron-like band centered at $k = \pi/a$, 
as shown in Fig. \ref{Fig:Bands}.  Due to the weak inter-orbital hopping, each of the bands 
is primarily derived from a single orbital, as indicated by the 
line thickness and colors in Fig. \ref{Fig:Bands}. One can therefore (loosely)  
regard the orbital character as an indicator of the band in this model. 
For example, the top most band is primarily composed of orbital $\gamma = 3$. The total bandwidth 
of the non-interacting model is $W = 4.9|t_{11}| = 2.45$ eV. This will serve as our unit of energy. 
We further set $a = 1$ as our unit of length. The interaction parameters are fixed 
to $U^\prime = U - 2J$, $J = U/4$, while $U$ is varied. This parameter regime results in a 
robust OSMP for intermediate values of $U$, which is our focus here. 

\subsection{DQMC and DMRG Calculations}
The model is studied using non-perturbative DQMC and DMRG methods. 
The details of these techniques can be found in Refs. \onlinecite{WhitePRB1989,DQMC_Review,RademakerPRB2013} (DQMC)  
and Refs. \onlinecite{White1} and \onlinecite{White2} (DMRG). 
These approaches are complementary to one another; 
DMRG works in the canonical ensemble and provides access to the ground state properties of the system while 
DQMC works in the grand canonical ensemble and provides access to finite temperatures and fluctuations in 
particle number. Both methods are capable of treating large cluster sizes such that non-local 
correlations can be captured without approximation for the specified Hamiltonian. 
 
The primary drawback to DQMC is the Fermion sign problem,\cite{Sign1,Sign2}  
which typically limits the range of accessible temperatures for many models. Indeed, when the 
spin-flip and pair hopping terms of the Hund's interaction are included in the Hamiltonian, we find that the 
model has a prohibitive sign problem.  But when these terms are 
neglected the corresponding sign problem becomes very mild,\cite{LiuPreprint2016} 
even in comparison to similar simplified multi-orbital models in 2D.\cite{RademakerPRB2013,BouadimPRB2008}  
Given that these terms do not qualitatively affect the 
phase diagram \cite{LiuPreprint2016} for the current model, we have neglected them here. 
This has allowed us to study clusters of up to $L = 24$ sites in length
($3L$ orbitals in total) down to temperatures as low as $\beta = 74/W$.\cite{LiuPreprint2016} 
At this low of a temperature we begin to see the onset of magnetic correlations in our cluster, 
however, as we will show, the OSMP forms at a much higher temperature. Since the latter phase is 
our focus here, we primarily show DQMC results for $\beta \le 19.6/W$ throughout. 
In all cases shown here, 
the average value of the Fermion sign is greater than $0.87 \pm 0.01$. Unless otherwise stated, all of our 
DQMC results were obtained on an $L =24$ site cluster with periodic boundary conditions and for 
an average filling of $\langle n\rangle = 4$ electrons, which corresponds to $2/3$ filling.

DQMC provides direct access to various quantities defined in the 
imaginary time $\tau$ or Matsubara frequency $\eye \omega_n$ axes.  
In Sec. \ref{Sec:Spectral} we will examine the spectral properties of our model, 
which requires an analytic continuation to the real frequency axis. 
This was accomplished using the method of Maximum Entropy,\cite{MaxEnt} 
as implemented in Ref. \onlinecite{FuchsPRE2010}.  

Our DMRG results were obtained on variable length chains with open boundary conditions. 
The chemical potential term in Eq. (\ref{Eq:H0}) is dropped for these calculations. 
In all of the DMRG calculations the truncation tolerance is between  $10^{-5}$ -- $10^{-7}$. 
We performed three to five full sweeps of finite DMRG algorithm and used 300 states 
for calculating both the ground state and the spectral function. 
Once the ground state is obtained using the standard DMRG algorithm, we 
computed the spectral function using the correction vector targeting 
in Krylov space  
\cite{DMRG_Method1,DMRG_Method2}, with an broadening of $\eta = 0.001$ eV. 

\section{Results}\label{Sec:Results} 
\subsection{Self-energies in the OSMP}
\begin{figure}[t]
 \includegraphics[width=0.66\columnwidth]{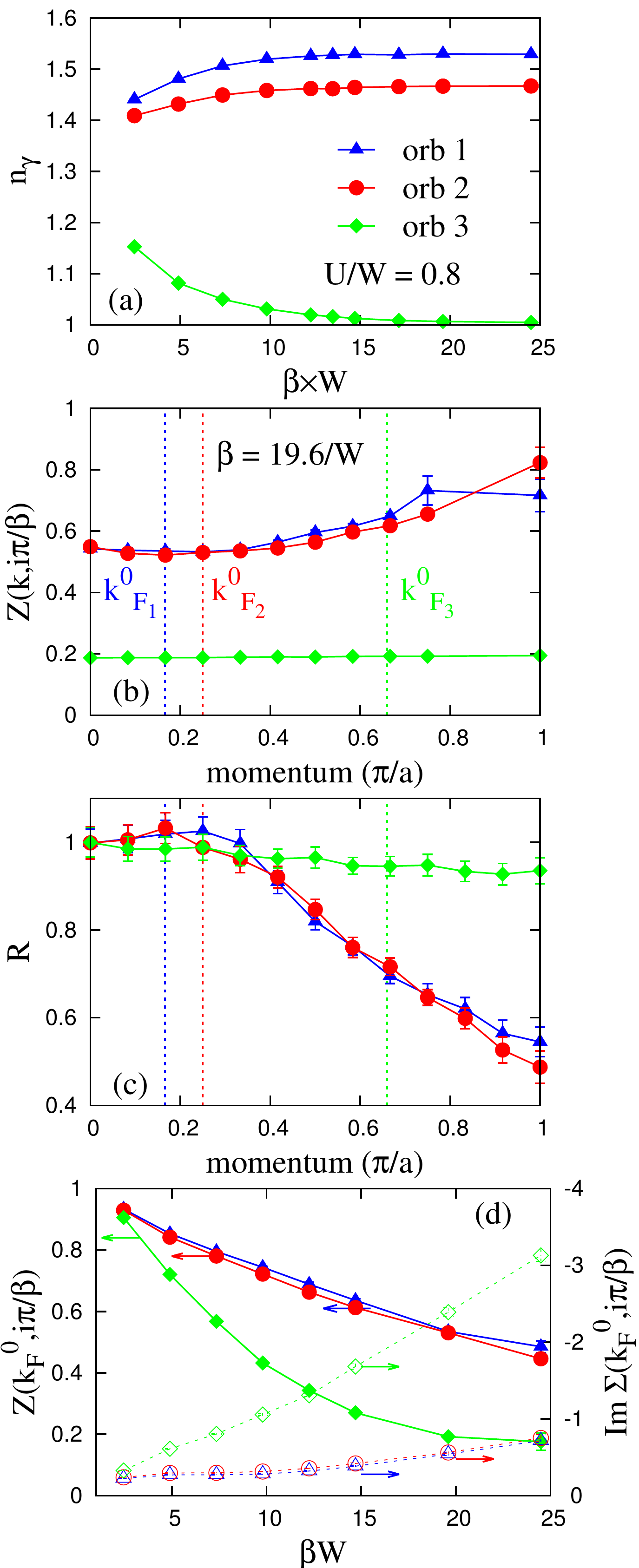}
 \caption{\label{Fig:n_beta} (color online) Orbitally resolved electronic
 properties for $U/W = 0.8$ ($W = 2.45$~eV) at
 different temperatures. (a) The temperature dependence of orbital occupations. 
 (b) The orbital resolved quasiparticle residue $Z_\gamma(k,\eye\pi/\beta)$ at an
 inverse temperature $\beta = 19.6/W$. (c)
 The normalized electron self energies Im$\Sigma_\gamma(k,\eye\pi/\beta)$ at 
 $\omega_n=\pi/\beta$ as a function of momentum. 
 Each curve is normalized by its $k = 0$ value to highlight the overall momentum dependence. 
 The scale is determined by $\mathrm{Im}\Sigma_\gamma(0,\eye\pi/\beta) = -0.53$, $-0.57$, 
 and $-2.53$ for $\gamma = 1,2,3$, respectively, and in units of the bandwidth $W$. The blue,
 red, and green dash lines in (b) and (c) correspond to the bare Fermi
 momentum of the non-interacting bands. Panel (d)
 shows orbitally resolved quasiparticle residues $Z_\gamma(k_\mathrm{F}^0,\eye\pi/\beta)$ and 
 self energies
 Im$\Sigma_\gamma(k_\mathrm{F}^0,\eye\pi/\beta)$ at Fermi momentum as a function of temperature. 
 In each panel, error bars smaller than the marker size have been
 suppressed for clarity.   
 }
\end{figure}

We begin by examining some of the standard metrics for the formation of an OSMP,
namely the average filling per orbital and the quasiparticle residue $Z_\gamma(k,\eye\omega_n)$.  
DQMC results for $\langle n \rangle = 4$ and $U/W = 0.8$ are summarized in Fig. \ref{Fig:n_beta}.  The temperature
dependence of the individual orbital occupations $\langle n_\gamma \rangle$,
plotted in Fig. \ref{Fig:n_beta}a, has the standard indications of the
formation of an OSMP: At high temperature (small $\beta$) we see noninteger fillings for all
three orbitals. As the temperature is lowered (large $\beta$), however, orbitals one and two
smoothly approach fillings of $\sim 1.53$ and $\sim 1.47$, respectively, while orbital three
locks into an integer value of exactly 1. In some studies these values for the 
average occupation are often taken as an
indication of an OSMP,\cite{RinconPRL2014,RinconPRB2014} where orbital three has undergone a transition to a Mott
insulating state while orbitals one and two host itinerant electrons. However,
as we will show, this is not always the case. For $U/W =
0.8$ the two fractionally filled orbitals are in fact itinerant, but for larger 
values of $U/W$ these same orbitals retain a fractional filling but are  
driven into an insulating state by the onset of orbital ordering in these two orbitals. 

The mixed itinerant/localized nature of the OSMP at $U/W = 0.8$ is reflected in the momentum 
dependence of quasi-particle residue $Z_\gamma(k,\eye\pi/\beta)$ and 
the orbitally resolved normalized self-energies 
$R(k) = {\mathrm{Im}}\Sigma_\gamma(k,\eye\pi/\beta)/{\mathrm{Im}}\Sigma_\gamma(0,\eye\pi/\beta)$, plotted in 
Figs. \ref{Fig:n_beta}c and \ref{Fig:n_beta}d, respectively, for $\omega_n = \pi/\beta$. 
The self-energy is extracted from the dressed Green's function using Dyson's equation 
\begin{equation}
\hat{G}^{-1}(k,\eye\omega_n) = \hat{G}^{-1}_0(k,\eye\omega_n) - \hat{\Sigma}(k,\eye\omega_n), 
\end{equation} 
where the $\hat{G}$ notation denotes a matrix in orbital space,  
$\hat{G}_0(k,\eye\omega_n) = [\eye\omega_n\hat{I} - \hat{H}_0(k)]^{-1}$ is the non-interacting Green's function, 
and $\hat{H}_0(k)$ is the Fourier transform of the non-interacting Hamiltonian defined in orbital space. 
The quasi-particle residue is obtained from the diagonal part of the 
self-energy using the identity 
\begin{equation}
\hat{Z}(k,\eye\pi/\beta) = \left(\hat{I} - \frac{\mathrm{Im}\hat{\Sigma}
(k,\eye\pi/\beta)}{\pi/\beta}\right)^{-1},
\end{equation} 
where $\hat{I}$ is a $3\times 3$ unit matrix.  
 
As can be seen from Fig. \ref{Fig:n_beta}c, the self-energies for each orbital have a sizable $k$-dependence 
at this temperature. (In this case we have normalized the self-energy by its value at $k = 0$ in 
order to highlight the overall momentum dependence. The magnitude of $\mathrm{Im}\Sigma_\gamma(0,\eye\pi/\beta)$ 
is given in the figure caption.) In the case of orbitals one and two, the magnitude of the self-energy 
varies by nearly 50\% throughout the Brillouin zone. In contrast, 
the momentum dependence of $\Sigma_3(k,\eye\pi/\beta)$ for orbital three is 
much weaker, varying by only 5-10\% and reflecting the localized nature  
of the carriers in these   
orbitals. Similarly, the quasi-particle residue for the orbital three is essentially 
momentum independent, while it increases for the two itinerant orbitals as 
$k$ tracks towards the zone boundary.  
The $k$ dependence at the remaining Matsubara frequencies accessible to our simulations 
(not shown) exhibits a similar trend, with orbitals one and two having a strong $k$-dependence 
while orbital three is nearly momentum independent at each $\omega_n$. 

The momentum dependence shown in Fig. \ref{Fig:n_beta} indicates that the local self-energy 
approximation introduced by DMFT may miss quantitative aspects of the electronic correlations 
in the OSMP with mixed itinerant and local characteristics. 
It should be noted that our results have been obtained in 1D, 
which is the worst case situation for DMFT. \cite{Akerlund_PRD}  
It is expected that the local approximation will perform better in higher dimensions, since DMFT 
becomes exact in the limit of infinite dimensions; however, it is unclear how well the method 
will capture similar non-local correlations in two dimensions relevant for the Fe-based superconductors. 
A recent study \cite{SemonPreprint} has argued that the local approximation is quite accurate 
for parameters relevant to the iron-based superconductors, however, it remains to be seen if this 
will remain true for all parameter regimes or when longer range fluctuations are included. 
Our results further highlight the need for the 
continued development of numerical methods capable of handling the strong 
Hubbard and Hund's interactions in intermediate dimensions and on extended clusters. 

Figure \ref{Fig:n_beta}d examines the temperature dependence of 
$Z(k^0_\mathrm{F},\frac{\eye\pi}{\beta})$ and Im$\Sigma(k^0_\mathrm{F},\frac{\eye\pi}{\beta})$ at the 
Fermi momenta $k^0_\mathrm{F}$ of the non-interacting system. 
(These are indicated by the dashed lines in \ref{Fig:n_beta}b and \ref{Fig:n_beta}c.) 
Here, we find indications of anomalous behavior for the itinerant electrons, 
where the quasiparticle residues of all three orbitals decrease with temperature.  
This is accompanied by an increase in Im$\Sigma(k_\mathrm{F},\frac{\eye\pi}{\beta})$ as $T$ 
is lowered. This is perhaps expected for orbital three, as $Z$ (Im$\Sigma$) for 
the localized orbitals should decrease (increase) as this orbital becomes more 
localized. For the itinerant orbitals, however, one would naively expect the 
self-energy to decrease as temperature is lowered, which is 
opposite to what is observed. We believe that this is due to the Hund's interaction between 
the itinerant electrons and the localized spins on 
orbital three. At this temperature we find no evidence of a magnetic ordering in our 
model,\cite{LiuPreprint2016} despite the fact that a local moment has clearly formed in 
the OSMP. 
This means that the orientation of the local moment is random and fluctuating 
at these temperatures. This produces a fluctuating potential 
acting on the itinerant electrons via the Hund's coupling, 
thus generating a residual scattering mechanism at low temperatures that reduces  
the quasiparticle residue and increases the self-energy.

\subsection{Momentum and Temperature Dependence of the Spectral Weight}
Next, we turn to the momentum dependence of the spectral weight for the three orbitals in the vicinity of the 
Fermi level. This can be estimated directly from 
the imaginary time Green's function, where the spectral weight at momentum $k$ is proportional to 
$\beta G(k,\tau=\beta/2)$.\cite{spectral} Using this relationship we do not have to 
perform the extra step of analytically continuing the data to the real frequency axis. 

\begin{figure}[t]
	\includegraphics[width=0.7\columnwidth]{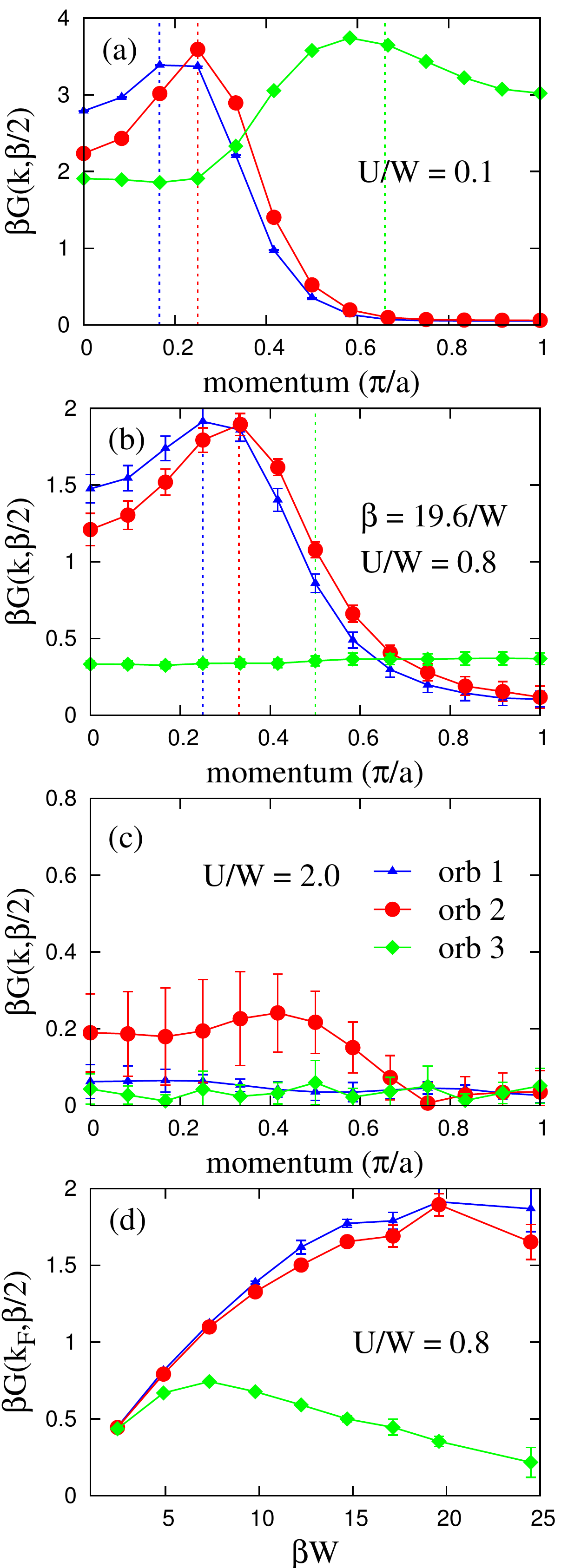}
        \caption{\label{Fig:Gbeta} (color online) The 
        momentum dependence of Green functions $G(k,\tau=\beta/2)$ for a) $U/W=0.1$,
        b) $0.8$, and c) $2.0$. The inverse temperature in all three cases is $\beta=19.6/W$. 
        The blue, red, and green dash lines in each panel indicate the Fermi
        momentum of the three non-interacting bands. 
        (d) $G(k_\mathrm{F},\tau=\beta/2)$ as a function of inverse temperatures $\beta$
        for the OSMP $U/W=0.8$. Error bars smaller than the marker size
        have been suppressed for clarity.  
	}
\end{figure}

Figures \ref{Fig:Gbeta}a-\ref{Fig:Gbeta}c summarize  
$\beta G(k,\beta/2)$ for $U/W=0.1$, $U/W=0.8$, and $U/W=2$, respectively. 
The results in the weak coupling limit ($U/W = 0.1$, Fig. \ref{Fig:Gbeta}a) are consistent with that 
of a fully itinerant system: all three orbitals have a maximal spectral weight at a 
momentum point very close to the Fermi momenta of the non-interacting system (indicated by the dashed lines). 
This is exactly the behavior one expects for a well-defined quasi-particle band dispersing 
through $E_\mathrm{F}$, where the peak in the spectral weight occurs at $k_\mathrm{F}$. The proximity of the 
peaks in $\beta G(k,\beta/2)$ to the non-interacting values of $k_\mathrm{F}$ indicates that the 
Fermi surface is only weakly shifted for this value of the interaction parameters. However, 
as we will show in Sec. \ref{Sec:BandShifts}, these shifts are band dependent. 

In the intermediate coupling regime ($U/W = 0.8$, Fig. \ref{Fig:Gbeta}b), where the
OSMP has formed, we again see both localized and itinerant characteristics. The
spectral weight of the localized orbital is small and independent of momentum,
as expected for the formation of a localized Mott state. Conversely, the
spectral weight of the remaining orbitals still exhibits a momentum dependence
characteristic of dispersive bands. Despite this, the total spectral weight is 
decreased, indicating that spectral weight has been transferred to higher binding 
energies by the Hubbard and Hund's interactions. 
This is also reflected in the position of the maximum spectral weight, which has  
shifted to a slightly larger $k$ value due to a 
renormalization of the Fermi surface by the interactions. We also observe
that the spectral weight at the zone boundary increases relative to the zone center, 
consistent with a flattening of the bands and a broadening of the spectral
function with increasing $U$. (This will be confirmed shortly when we examine
the spectral functions directly.) A similar transfer of spectral weight was observed in a 
two-dimensional cluster DMFT study.\cite{NomuraPRB2015} 

The temperature evolution of spectral weight $\beta G(k_\mathrm{F},\beta/2)$ at the 
Fermi momentum for the OSMP ($U/W = 0.8$) is shown in Figure \ref{Fig:Gbeta}d. 
In a metallic system one generally expects the spectral weight at the Fermi
level to increase as the temperature is decreased. 
Initially, this is what is observed for all three orbitals, 
however, the spectral weight for orbital three reaches a maximum around 
$\beta = 7.5/W$ before decreasing as the temperature is lowered further and the OSMP gap 
forms on this orbital. Conversely, the spectral weight of the itinerant orbitals 
continues to rise until saturating at $\beta/W \approx 15$. This saturation is 
again due to the presence of a residual scattering channel, which we associate with 
the fluctuating localized spins present on the localized orbital three. 

The $U/W = 0.8$ results confirm the mixed itinerant/local character of the 
model at intermediate coupling.  
When the value of $U$ is further increased, we find that all three bands 
become localized while maintaining partial occupancies for each band.  
To demonstrate this, Fig. \ref{Fig:Gbeta}c shows results for $U/W = 2$.  
In this case, the orbital occupations 
for the three orbitals are $\langle n_1 \rangle = 1.55$, $\langle n_2 \rangle = 1.44$, 
$\langle n_3 \rangle = 1$, which are similar to those obtained   
at $U/W = 0.8$. At face value one might therefore conclude that the system is in an 
OSMP,\cite{RinconPRL2014,LiuPreprint2016} however, an examination of the spectral 
weight reveals that the system is in fact insulating. 
As can be seen in Fig. \ref{Fig:Gbeta}c, at $U/W = 2$  and $\beta = 19.6/W$, 
$\beta G(k,\beta/2)$ is nearly momentum independent and the total spectral weight of 
all three orbitals has significantly decreased (note the change in scale of the y-axis). This 
behavior is indicative of the formation of a charge gap throughout the 
Brillouin zone. The ultimate origin of this insulating behavior is the formation 
of a long-range orbital ordering, as we will show in Sec. \ref{Sec:Spectral}. 

\subsection{Band-dependent Fermi surface renormalization}\label{Sec:BandShifts}
\begin{figure}[t]
 \includegraphics[width=0.7\columnwidth]{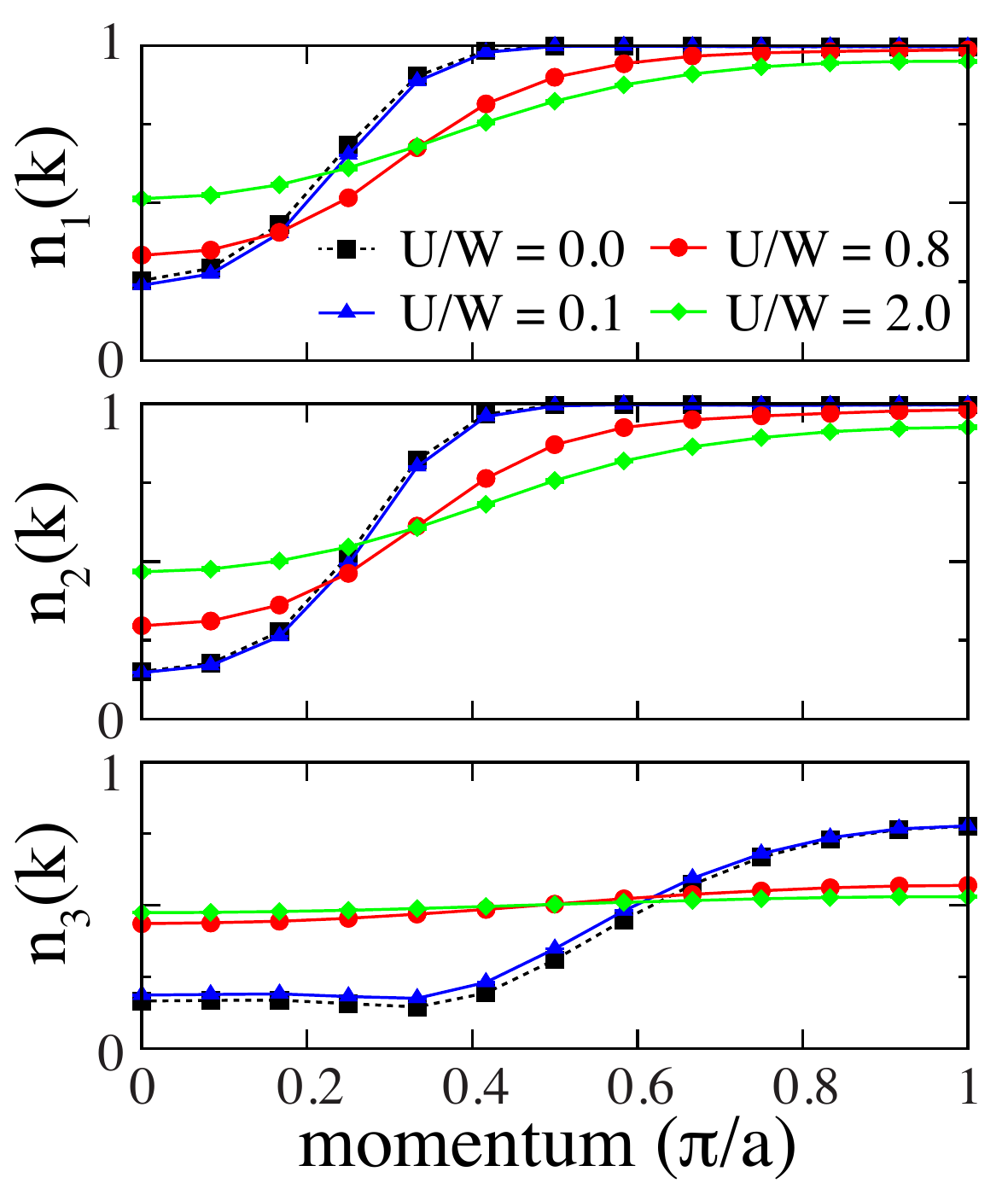}
 \caption{\label{Fig:n_k} (color online) 
 The momentum dependence of the number operator 
 $n_\gamma(k) = \frac{1}{2}\sum_\sigma \langle c^\dagger_{\bk,\gamma,\sigma} c_{\bk,\gamma,\sigma}\rangle$ for each 
 band. Results are shown for the non-interacting case $U = 0$ (black dashed, $\square$), 
 $U/W = 0.1$ (blue solid, $\bigtriangleup$), $U/W = 0.8$ (red solid $\circ$), and 
 $U/W = 2$ (green solid $\diamond$) and at an inverse temperature of $\beta = 19.6/W$.}
\end{figure}

It is now well known that {\em ab initio} band structure calculations based on 
density functional theory (DFT) do not describe 
the electronic structure of the iron based superconductors as measured in ARPES 
experiments. (For a recent review, see Ref. \onlinecite{vanRoekeghem}.)  Generally speaking, 
the calculated band structure usually needs to be rescaled by an overall factor, which 
is attributed to reduction in bandwidth driven by electronic correlations. In addition, 
the size of the Fermi surfaces is often overestimated by DFT in comparison to 
measurements. A prominent example of this is LiFeAs,\cite{Chi} where the inner most hole pocket 
realized in nature is 
substantially smaller than the one predicted by DFT \cite{Klaus,Klaus2}. In order to correct this, 
the electron- and hole-bands need to be shifted apart,\cite{vanRoekeghem} which 
requires a momentum-dependent self-energy correction. 

We examine this issue within our model in Fig. \ref{Fig:n_k}, which plots the 
expectation value of the orbitally-resolved number operator in momentum space 
$n^\pdag_\gamma(k) = \frac{1}{2}\sum_\sigma \langle c^\dagger_{\bk,\gamma,\sigma} c^\pdag_{\bk,\gamma,\sigma}\rangle$
for various values of the interaction strength. 
In the non-interacting limit, and in a single-band case, this quantity is equal to the 
Fermi-Dirac distribution and the location of the leading edge corresponds to $k_\mathrm{F}$. 
In a multi-band system the mixing of the orbital character complicates this 
picture; however, in our model the leading edge still corresponds to $k_\mathrm{F}$ 
due to the weak hybridization between orbitals. In the weak coupling case ($U/W = 0.1$) 
we observe a small shift in the position of the leading edge. Within error bars, the 
curve $n_1(k)$ and $n_2(k)$ shift to slightly larger momenta while 
$n_3(k)$ shifts towards smaller momenta. This indicates that the size of the 
Fermi surfaces are increasing and the electron-like and hole-like bands are  
shifted towards one another by the interactions. This trend continues as $U/W$ is 
increased to $0.8$; however, in this case the electron-like band is significantly 
smeared out due to the formation of the OSMP. 

We note that the direction of the band shifts is reversed from what is generally 
required for the two-dimensional iron-based superconductors, where the calculated 
hole-like Fermi surfaces generally need to be shrunk relative to the electron-like 
Fermi surfaces.  We attribute this to differences in the underlying tight-binding model 
and differences in dimensionality. In this light, it would be interesting to compare the ARPES 
observed band structures in the quasi-one-dimensional pnictides against the predictions 
of our model and DFT calculations.\cite{Patel} Nevertheless, our results do show that non-local 
correlations arising from a local interaction  
can produce relative shifts of the electron-like and hole-like bands 
in a multi-orbital system.  

\subsection{Spectral Properties}\label{Sec:Spectral}
\subsubsection{Intermediate Coupling $U/W = 0.8$}
\begin{figure}[t]
	\includegraphics[width=0.75\columnwidth]{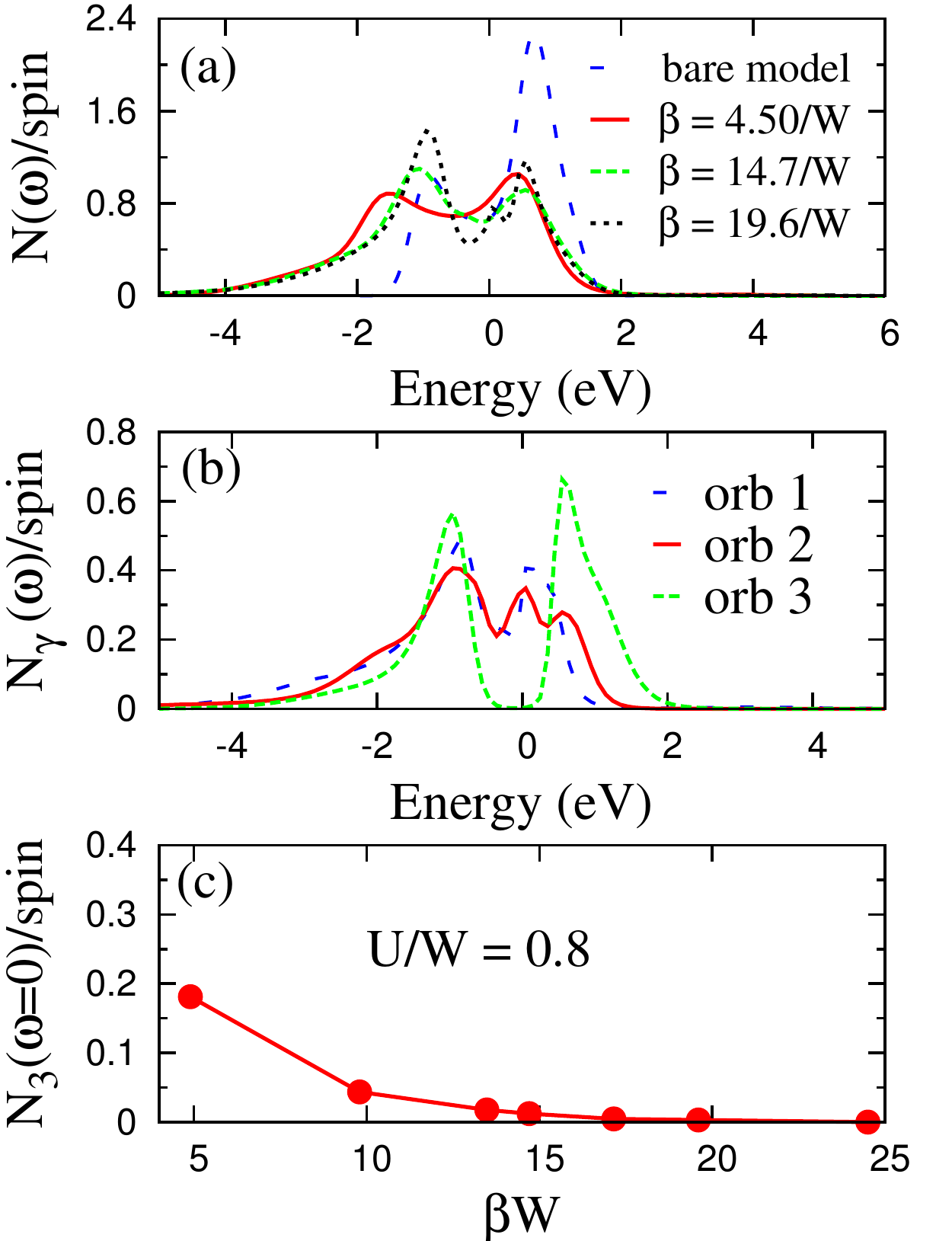}
        \caption{\label{Fig:dos} (color online)  (a) The density of states at
different temperatures. (b) The orbitally-resolved density of states for each
orbital at an inverse temperature $\beta=19.6/W$. (c) The density of states at
the Fermi surface of the orbital 3 as a function of inverse temperatures
$\beta$. The Coulomb interaction strength is $U/W=0.8$ in all three graphs. 
	}
\end{figure}

We now examine the spectral properties of the model, beginning with
the OSMP.  Figure \ref{Fig:dos}a shows the temperature evolution of the total
density of states (DOS) at $U/W = 0.8$, which is obtained from the trace of the
orbital-resolved spectral function $N(\omega) = \sum_{k,\gamma}
-\frac{1}{\pi}{\mathrm{Im}}\hat{G}_{\gamma\gamma}(k,\omega+i\delta)$.  In the
non-interacting limit (the long-dashed (blue) curve), the DOS has a double peak
structure, where the lower (upper) peak corresponds to the bands derived from
orbitals one and two (orbital three). The overall structure of the DOS in the
interacting case is similar at high temperatures, but some spectral weight is
transferred to a broad incoherent tail extending to lower energies.  As the temperature is
decreased, the peak on the occupied side shifts towards the Fermi level and 
sharpens. At the same time, a small amount of spectral weight is  
transferred from the vicinity of the Fermi level into this peak. The appearance 
of this apparent ``pseudogap" is a direct consequence of the OSMP
forming on orbital three, which is easily confirmed by examining the
orbital-resolved DOS $N_\gamma(\omega) = -\frac{1}{\pi}\sum_k
\mathrm{Im}\hat{G}_{\gamma,\gamma}(k,\omega)$ shown in Fig. \ref{Fig:dos}b.
As can be clearly seen, orbitals one and two have a finite DOS at $\omega = 0$,
while orbital three is fully gapped at low-temperature. 

\begin{figure*}[t]
	\includegraphics[width=0.8\textwidth]{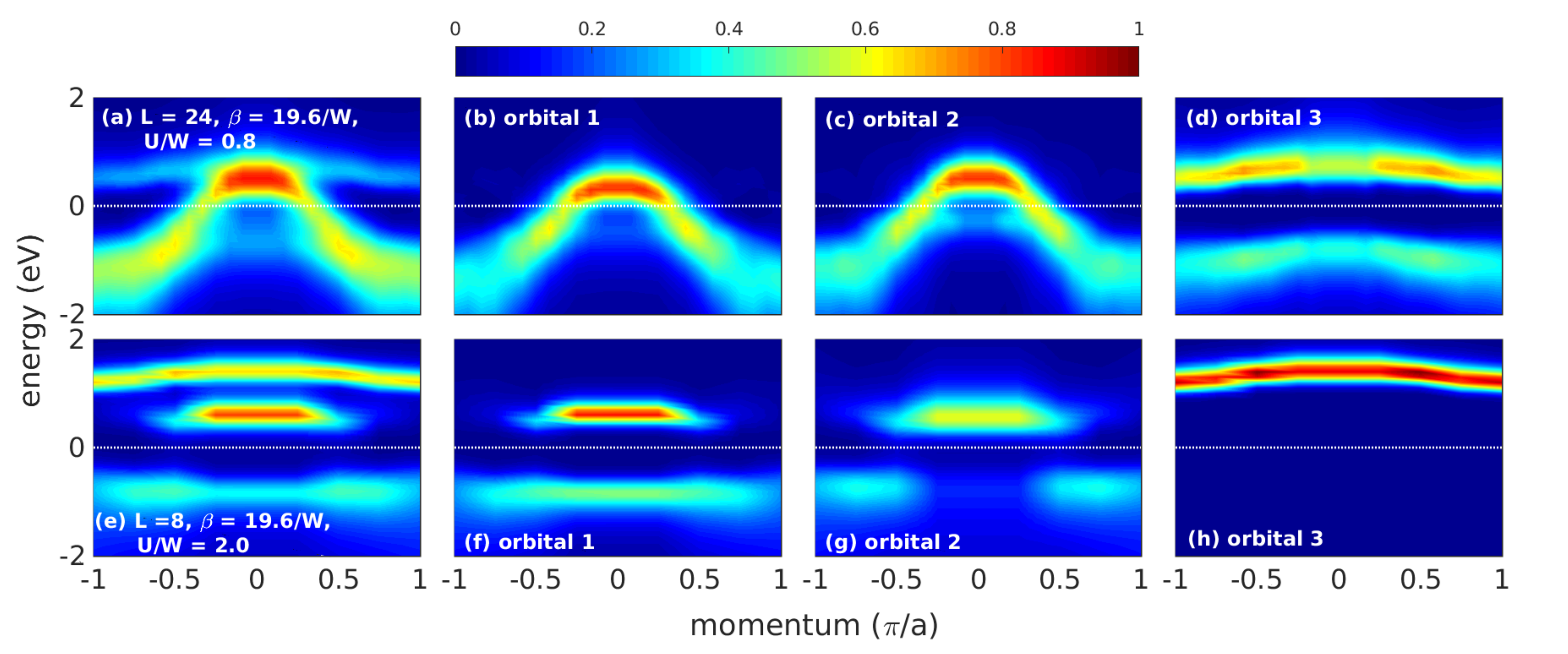}
        \caption{\label{Fig:spectral} (color online) (a) The spectral function
for $U/W=0.8$. (b), (c), and (e) are the orbital 1, 2, and 3 parts of the
spectral function in (a), respectively. (e) The spectral function for $U/W =2$.
(f), (g), and (h) are the orbital 1, 2, and 3 parts of the spectral function in
(e), respectively. The dash white line labels the Fermi surface. The inverse
temperature is set as $\beta=19.6/W$. Results where obtained with Maximum Entropy DQMC.
	}
\end{figure*}

We also begin 
to see the formation of an additional peak near the Fermi level 
at the lowest temperature we examined ($\beta = 19.6/W$).
This feature is more clearly seen in the orbital-resolved DOS (Fig. \ref{Fig:dos}b), 
where it is found to originate from the itinerant orbitals. 
This peak is due to a hybridization between the itinerant 
and localized orbitals, which is observable in the $k$-resolved spectral functions 
(see Fig. \ref{Fig:spectral}).

The relevant temperature scale for the formation of the OSMP can be estimated
by tracking $N_{3}(0)$ as a function of
temperature, as shown in Fig. \ref{Fig:dos}c. Here, a continuous suppression of
$N_3(0)$ is observed, with the value reaching zero at $\beta \approx 20/W$.  The
rate at which $N_3(0)$ decreases also undergoes a distinct change at $\beta 
\approx 10/W$, which coincides with the temperature at which the spectral
weight for this orbital at $k_\mathrm{F}$ is largest (see Fig.
\ref{Fig:Gbeta}d). We interpret this to mean that the Mott gap on orbital three
begins to form at $\beta W \approx 10$ (on the $L = 24$ site lattice), growing
continuously from zero as the temperature is lowered.  In this case, the finite
spectral weight between $\beta W = 10$ -- $20$ is due to thermal
broadening across this gap. Since we have observed similar behavior on 
smaller clusters with DQMC and at zero temperature using DMRG, we 
believe that the transition to the OSMP will survive 
in the thermodynamic limit, however, the gap magnitude has some finite size 
dependence. 

The extended length of our 1D cluster grants us access to the momentum 
dependence of the spectral function, which is shown in Fig. \ref{Fig:spectral}. 
The top row of Fig. \ref{Fig:spectral} shows the 
results in the OSMP with $U/W = 0.8$ and $\beta = 19.6/W$, which is the same 
parameter set used in Fig. \ref{Fig:dos}.  
The total spectral function $A(k,\omega) =-\frac{1}{\pi}\mathrm{Tr}\left[\mathrm{Im}\hat{G}(k,\omega)\right]$ 
is shown in Fig. \ref{Fig:spectral}a and the orbital-resolved components 
$A_\gamma(k,\omega)=-\frac{1}{\pi}\mathrm{Im}\hat{G}_{\gamma\gamma}(k,\omega)$ 
are shown in Figs. \ref{Fig:spectral}b-d, as indicated. 
The lower row of Fig. \ref{Fig:spectral} shows similar results obtained for 
$U/W = 2$ and $L = 8$. (In this case a smaller cluster is sufficient 
due to the non-dispersing nature of the band dispersions.)

The results in the OSMP with $U/W = 0.8$ reveal localized and itinerant 
characteristics that are consistent with the spectral weight analysis 
presented earlier. The itinerant orbitals primarily contribute to 
dispersing bands that track through the $E_\mathrm{F}$ ($\omega = 0$), while 
orbital three has split into two relatively dispersionless upper and lower Hubbard bands  
above and below $E_\mathrm{F}$. At first glance, these Hubbard bands appear to be 
sharper than the corresponding Hubbard bands in the single-band Hubbard model; however, 
an examination of the DOS (Fig. \ref{Fig:dos}b) reveals that they are spread out over 
an energy interval that is larger than the non-interacting bandwidth of the 
top most band ($W_3 \sim 0.3W \sim 0.735$ eV). 
In addition to the formation of the Hubbard bands for orbital 3, we also observe two additional 
effects. The first is an expected narrowing of the bandwidth of the 
itinerant bands. For this parameter 
set we obtain $W_1 \sim 1.7$ and $W_2 \sim 1.65$ eV for orbitals one and two, 
respectively, which should be compared to the non-interacting values of 1.88 and 1.97 eV.   
The second is the aforementioned hybridization and level repulsion between the itinerant and 
localized orbitals. This is manifest in the spectral 
function as a slight ``buckling" of orbital three's upper Hubbard band near $k = 0$, 
and the tracking orbital one's spectral weight along $E_\mathrm{F}$ near $k = \pm\pi/2a$.  
It is this trailing intensity that forms the peak observed in the DOS 
just above the Fermi level at low temperatures.

\subsubsection{Strong Coupling $U/W = 2$}
The spectral properties of the model are very different when 
the Hubbard interaction is increased to $U/W = 2$. 
In this case, the total spectral function (Fig. \ref{Fig:spectral}e) and its 
orbitally-resolved components (Fig. \ref{Fig:spectral}f-\ref{Fig:spectral}h) 
all split into relatively flat Hubbard-like bands above and below $E_\mathrm{F}$. 
(In the case of orbital three, the lower band below $E_\mathrm{F}$ has been pushed outside of 
the energy range shown in the figure.) For this value of the interaction strength there is 
no spectral weight at the Fermi level, and the system is insulating even though 
orbitals one and two have on average 1.55 and 1.44 electrons/orbital, 
respectively. 
(These values are obtained both from the measured equal time orbital occupancies, and 
from integrating the total spectral weight above and below $E_\mathrm{F}$.)  

The imaginary axis spectral weight analysis (Fig. \ref{Fig:Gbeta}c) and the spectral function 
analysis (Fig. \ref{Fig:spectral}) both indicate that for 
$U/W = 2$ the model is an insulator. The origin of this behavior 
is the combined action of the Hund's coupling and the onset of an orbital ordering 
of the itinerant orbitals. All indications show that orbital three 
has already undergone an orbital selective Mott phase transition (OSMT) 
when $U/W = 2$. This has the effect of localizing one electron per site within this 
subset of orbitals while leaving 
three additional electrons to be distributed among the remaining two itinerant orbitals. 
A sizable Hund's coupling will decouple the 
individual orbitals when the crystal field splittings are smaller 
than the bandwidth of the material.\cite{deMediciPRL2014} This is precisely the situation at hand, 
and thus the remaining nominally itinerant orbitals are decoupled from the localized orbital
by the large $J = U/4$. This results in an effective nearly-degenerate two-band system 
with (nearly) three-quarters filling. This is special case for the two-orbital 
Hubbard model, which is prone to orbital ordering 
in one and two-dimensions.\cite{HeldEPJB,ChanPRB2009,KuboPRB2002}  

\begin{figure}
 \includegraphics[width=0.9\columnwidth]{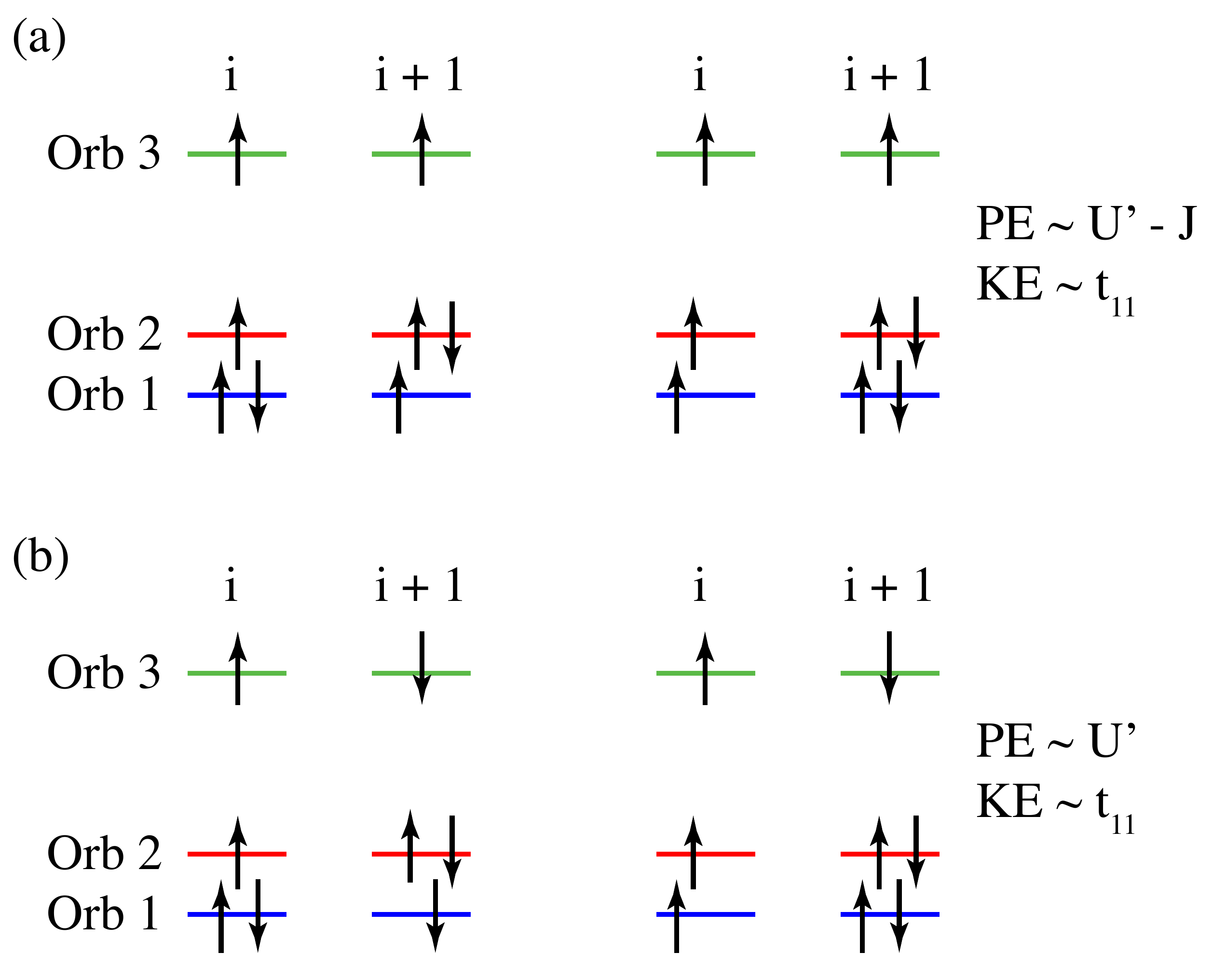}
 \caption{\label{Fig:OrbitalOrdering}  
 A cartoon sketch of the relevant charge fluctuation processes leading to 
 the insulating state when $U/W = 2$ assuming (a) ferromagnetic and (b) 
 antiferromagnetic nearest neighbor correlations within the orbital that has 
 undergone the orbital selective Mott transition (orbital three). 
 }
\end{figure}

The situation is sketched in Fig. \ref{Fig:OrbitalOrdering}. 
Assuming ferromagnetic nearest neighbor correlations for orbital three, 
we have a low-energy ground state configuration as shown in the left side of \ref{Fig:OrbitalOrdering}a. 
Here, orbitals one and two adopt alternating double occupations 
in order to maximize their delocalization energy through virtual hopping 
processes. This results in near-neighbor orbital correlations. 
Subsequent charge fluctuations such as the one shown 
in the right side of the Fig. \ref{Fig:OrbitalOrdering}a cost a potential energy 
$PE \sim U^\prime - J = W/2$. This is compensated for by a kinetic energy gain 
$KE \sim 4t_{11} \sim 4W/4.9$. The ratio between these competing energy 
scales is $\sim 5/8$, suggesting that charge fluctuations are strongly 
suppressed by the strong electronic correlations in this subsystem. 
Note that the situation is worse for antiferromagnetic nearest neighbor correlations 
in orbital three. The energy cost in this case 
increases to $\sim U^\prime$, as shown in Fig. \ref{Fig:OrbitalOrdering}b. 
Thus both ferro- and antiferromagnetic correlations in orbital three will 
suppress charge fluctuations and promote orbital ordering. 
Since the type of magnetic correlations does not matter, such orbital ordering tendencies 
can be expected in the paramagnetic phases, provided the localized moments 
have formed in orbital three.  This picture is then consistent with 
insulating behavior (and short-range orbital ordering tendencies, see below)
at high temperatures, where no magnetic correlations are observed. 

\begin{figure}
 \includegraphics[width=0.8\columnwidth]{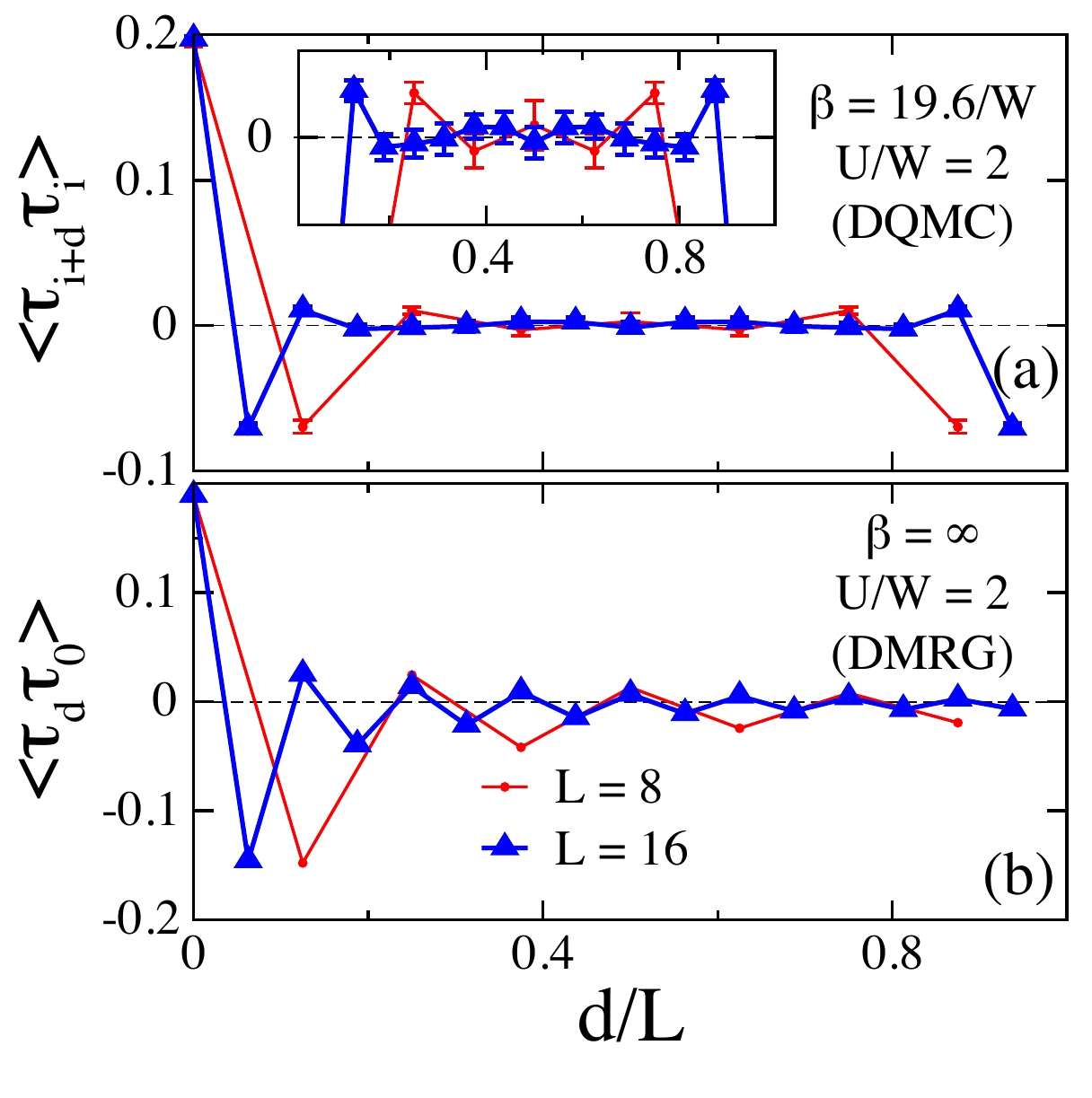}
 \caption{\label{Fig:Orbital_correlation} 
 Results for the orbital correlation function 
 for the system in the strong coupling case $U/W = 2$. Results are obtained 
 at (a) finite temperature using DQMC and (b) $T = 0$ ($\beta = \infty$) 
 using DMRG. 
 In both cases, results are shown on $L = 8$ (red dots) and $L = 16$ 
 (blue triangles) chains. The DQMC results were obtained on a chain with 
 periodic boundary conditions. The DMRG results were obtained on a chain with  
 open boundary conditions. 
 }
\end{figure}

\begin{figure}
 \includegraphics[width=0.8\columnwidth]{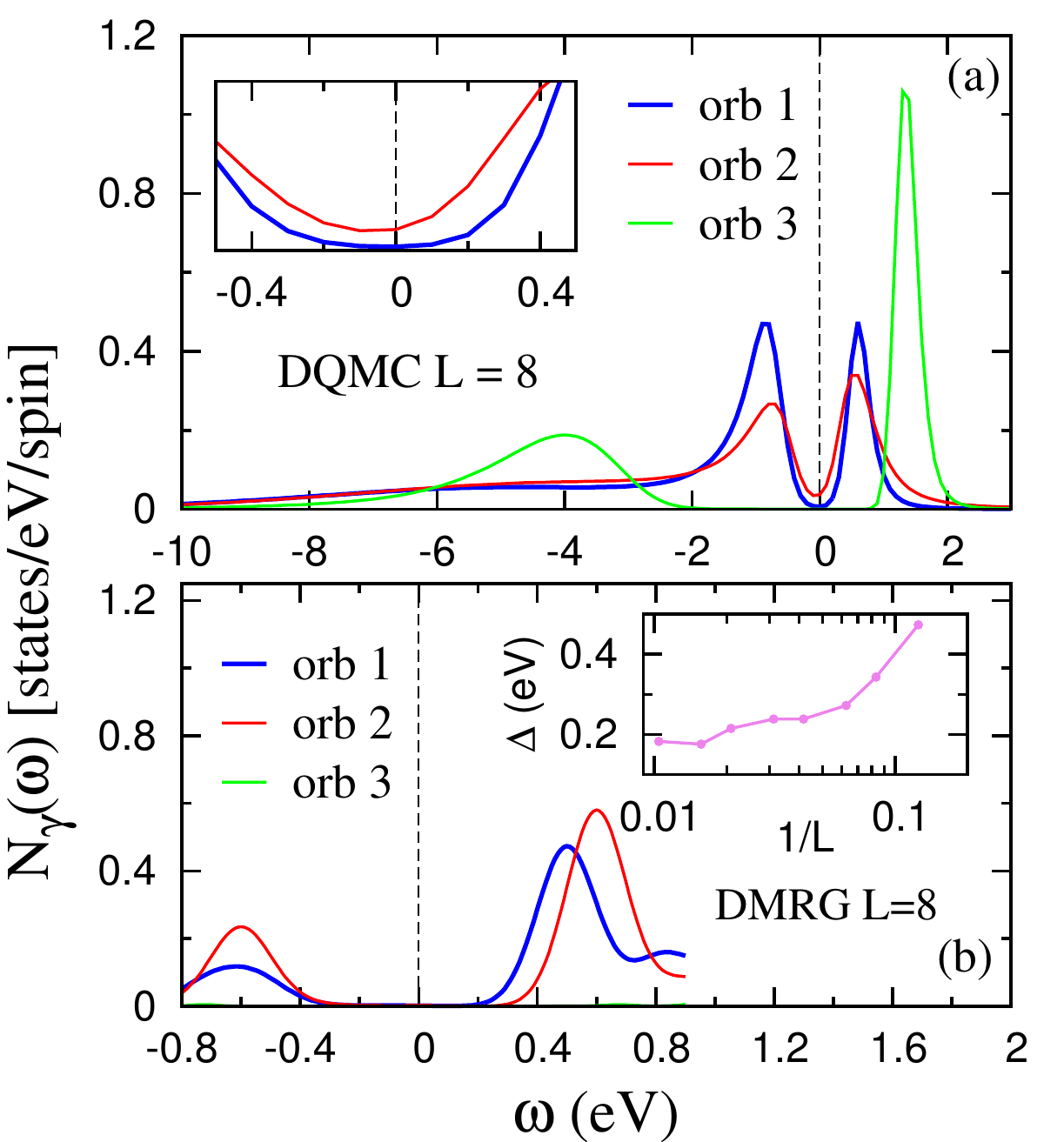}
 \caption{\label{Fig:Ueq2} 
 Results for the orbitally-resolved density of states for each orbital 
 obtained for $U/W = 2$ and on $L = 8$ site 
 chains. Panel (a) shows DQMC results at $\beta = 19.6/W$ 
 and the inset zooms in to energy around Fermi
 surface. Panel (b) shows DMRG results for the same conditons but at 
 zero temperature ($\beta = \infty$). The inset plots a finite size 
 scaling analysis of the charge gap obtained within DMRG (see text). 
 The dash line in both panels indicates the Fermi energy.
 }
\end{figure}

We verify this picture explicitly in Fig. \ref{Fig:Orbital_correlation}, which plots the 
equal-time orbital correlation function $\langle \hat{\tau}_{i+d}\hat{\tau}_i\rangle$, 
with $\hat{\tau}_{i} = (\hat{n}_{i,2}-\hat{n}_{i,1})$. Here, results are shown for finite 
temperature DQMC calculations (Fig. \ref{Fig:Orbital_correlation}a) and zero 
temperature DMRG calculations (Fig. \ref{Fig:Orbital_correlation}b) and with $U/W=2$ in 
both cases. The ``long-range" (with respect to the cluster size) 
anti-ferro-orbital correlation is clear in the  
zero temperature results obtained on $L = 8$ and $L=16$ chains. 
At finite temperatures ($\beta = 19.6/W$) we find that the 
orbital correlations are suppressed at long distances, but local 
anti-ferro-orbital correlation remains on shorter length scales. These 
combined results demonstrate the 
presence of short-range orbital correlations at higher temperatures, which 
grow in length as the temperature is decreased.   
The corresponding orbitally resolved DOS are plotted in Fig. \ref{Fig:Ueq2} 
for both cases. Both methods predict that the system is insulating, with a charge 
gap width on orbitals one and two of about 0.5 eV. The presence of a gap at 
finite temperature also confirms that the short range orbital correlations 
are sufficient to open a gap in the spectral function. Finally, we stress these 
results will survive in the thermodynamic limit $L \rightarrow \infty$. 
This is confirmed in the inset in Fig. \ref{Fig:Ueq2}b, which shows 
the evolution of the $T = 0$ gap $\Delta$ as a function of $L$, 
as obtained from DMRG. Here, the gap size decreases with 
increasing chain lengths, however, it saturates to 0.2 eV for an infinite length 
chain. %This strong size effect indicates a long correlation length in 
%electron-electron interactions. 

\section{Discussion and Summary}\label{Sec:Discussion}

We have performed a momentum-resolved study of a multi-orbital model defined on 
extended 1D chains using non-perturbative DQMC and DMRG. 
This has allowed us to compute the 
several properties of an OSMP in a momentum resolved manner without resorting to 
approximate methods. We find that several properties do indeed exhibit significant momentum 
dependencies, not be captured by local approximations 
introduced by DMFT; however, the 1D case we have considered 
represents the worst case for DMFT. In that sense our results 
complement existing DMFT efforts by providing analysis in a region where the method is expected to 
perform badly.   

Our results establish the hierarchy of charge and magnetic orderings in this model.  
At low temperatures, our DMRG calculations (as well as those in Ref. \onlinecite{RinconPRL2014}) 
demonstrate that orbital three is ferromagnetically ordered at $T = 0$. Contrary to this, 
our finite temperature DQMC calculations find no indications of any magnetic order 
for $\beta < 19.6/W$; the magnetic structure factor $S(q)$ is completely featureless 
as a function of $q$ at these temperatures. Despite this, our finite $T$ calculations find an orbital-selective 
Mott phase, as well as a fully insulating phase arising due to short-range orbital ordering, depending on 
the strength of the Hubbard interaction $U$.   
We therefore conclude that the charge ordering occurs before any magnetic ordering in this model. 

The results shown in Fig. \ref{Fig:n_beta}d and  \ref{Fig:Gbeta}d 
show  that orbital three in our model, which has the narrowest band width, undergoes a transition to 
a Mott phase at $\beta W \sim 10-15$. This in combination with the lack of magnetic signal 
means that OSMP in this parameter regime is a true Mott phase as opposed to a Slater insulator where 
the insulating behavior is driven by magnetism. 
Our results also demonstrate that it is insufficient to identify an OSMP using 
the orbital occupations only in some instances. One should be particularly 
careful in regions of parameter space where the itinerant bands have average occupations 
close to special cases known for one and two-orbital Hubbard models. In our case, 
the average fillings of the itinerant orbitals are $\langle n_1\rangle \sim 1.53$ 
and $\langle n_2 \rangle \sim 1.47$, values very close to the special case of $3/4$ filling in 
a degenerate two-band Hubbard model. At zero temperature, our DMRG results obtain 
fillings of 1.5 for each orbital. 

Finally, we discuss our results in the context of recent experimental work. 
ARPES results for AFe$_2$As$_2$ have found evidence that the OSMP in these 
materials disappears as the temperature is lowered.\cite{YiPRL2013} This 
behavior was explained using a slave-boson approach and attributed to an increase in 
entropy associated with the OSMP. Our results do not show this behavior, and the 
OSMP is found at low temperature as one might naively expect. 
This difference may be 
related to the differences in the dimensionality (one vs. two) or number of orbitals 
(three vs. five) between the models or the differences between our non-perturbative 
approach and other mean-field methods. However, one would expect the entropy to be 
more important in one dimension. This highlights the need for  
continued application of non-perturbative methods to tractable multi-orbital Hubbard 
models.  

{\em Acknowledgements} --- 
The authors thank G. Liu for useful discussions. 
S. L., Y. W., and S. J. are supported by the
University of Tennessee's Science Alliance Joint Directed Research and
Development (JDRD) program, a collaboration with Oak Ridge National Laboratory.
N. K. and E. D. were  supported  by  the  National  Science  Foundation  (NSF) under 
Grant No. DMR-1404375. 
Y. T. and T. A. M. acknowledge support by the Laboratory Directed Research
and Development Program of Oak Ridge National Laboratory, managed by
UT-Battelle, LLC, for the U.S. Department of Energy. 
Part of this work was conducted at the Center for Nanophase Materials Sciences, 
sponsored by the Scientic User Facilities Division (SUFD), BES, DOE, 
under contract with UT-Battelle. A.N. and G.A. acknowledge support by
the Early Career Research program, SUFD, BES, DOE.  CPU time was 
provided in part by resources supported by the University of Tennessee and Oak
Ridge National Laboratory Joint Institute for Computational Sciences
(http://www.jics.utk.edu).

\end{document}